\documentclass[aps,superscriptaddress,tightenlines]{revtex4}
\usepackage{graphicx,amssymb,amsbsy,bm,amsmath}

\newcommand{\vx}{\ensuremath{\vec{x}}}
\newcommand{\vk}{\ensuremath{\vec{k}}}

\newcommand{\be}{\begin{equation}}
\newcommand{\ee}{\end{equation}}
\newcommand{\bea}{\begin{eqnarray}}
\newcommand{\eea}{\end{eqnarray}}
\begin{document}
\title{Particle decay in inflationary cosmology}
\author{D. Boyanovsky}
\email{boyan@pitt.edu} \affiliation{Department of Physics and
Astronomy, University of Pittsburgh, Pittsburgh, Pennsylvania
15260, USA}
\affiliation{Observatoire de Paris, LERMA.
Laboratoire Associ\'e au CNRS UMR 8112.
 \\61, Avenue de l'Observatoire, 75014 Paris, France.}
\affiliation{LPTHE, Universit\'e
Pierre et Marie Curie (Paris VI) et Denis Diderot (Paris VII),
Tour 16, 1er. \'etage, 4, Place Jussieu, 75252 Paris, Cedex 05,
France}
\author{H. J. de Vega}
\email{devega@lpthe.jussieu.fr} \affiliation{LPTHE, Universit\'e
Pierre et Marie Curie (Paris VI) et Denis Diderot (Paris VII),
Tour 16, 1er. \'etage, 4, Place Jussieu, 75252 Paris, Cedex 05,
France}\affiliation{Observatoire de Paris, LERMA.
Laboratoire Associ\'e au CNRS UMR 8112.
 \\61, Avenue de l'Observatoire, 75014 Paris, France.}
\affiliation{Department of Physics and Astronomy,
University of Pittsburgh, Pittsburgh, Pennsylvania 15260, USA}
\date{\today}
\begin{abstract}
We investigate the relaxation and decay of a particle during inflation by
implementing the dynamical renormalization group. This investigation
allows us to give a meaningful definition for the decay rate in an expanding
universe. As a prelude to a more general scenario,
the method is applied here to
study the decay of a particle in de Sitter inflation via a trilinear coupling
 to massless conformally coupled particles, both for wavelengths
 much larger and much smaller than the Hubble radius.
For superhorizon modes we find that the decay is of the form
 $\eta^{\Gamma_1}$ with $\eta$ being conformal time and we give an
 explicit expression for $\Gamma_1$ to leading order in the coupling
which has a noteworthy interpretation in terms of the Hawking temperature
 of de Sitter space-time. We show that if the mass $M$ of the
decaying field is $\ll  H$ then  the decay rate during inflation
is \emph{enhanced}  over the Minkowski space-time result by a factor
 $2H/\pi M$.  For wavelengths much smaller
 than the Hubble radius we find that the decay law is
 $e^{-\frac{\alpha}{kH}  C(\eta)}$ with $C(\eta)$ the scale factor and
$\alpha$ determined
 by the strength of  the trilinear coupling.  In all cases we find
 a substantial enhancement in the decay law as compared to
 Minkowski space-time. These results suggest potential implications for the
 spectrum of scalar density fluctuations as well as  non-gaussianities.
\end{abstract}


\maketitle

\section{Introduction}

Inflation was originally proposed to solve several outstanding
problems of the standard Big Bang model
\cite{guth,kolb,coles,liddle,riottorev} thus becoming an
important paradigm in cosmology. At the same time that inflation
solves these problems it also  provides a natural mechanism for
the generation of scalar density fluctuations that seed large
scale structure, thus explaining the origin of the temperature
anisotropy in the cosmic microwave background (CMB),  as well as
tensor perturbations (primordial gravitational waves). Recently
the Wilkinson Microwave Anisotropy Probe (WMAP) collaboration has
provided a full-sky map of the temperature fluctuations of the
cosmic microwave background (CMB) with unprecedented accuracy and
an exhaustive analysis of the data confirming the basic and robust
predictions of inflation\cite{WMAP,WMAP1,WMAP2}.

During inflation quantum vacuum fluctuations are generated with
physical wavelengths that grow faster than the Hubble radius, when
the wavelength of these perturbations crosses the horizon these
perturbations freeze out and decouple\cite{kolb,liddle,riottorev}.
Wavelengths that are of cosmological relevance today  re-enter the
horizon during matter domination when the scalar (curvature)
perturbations induce temperature anisotropies that are imprinted
on the CMB at the last scattering surface\cite{hu,hu2}. Generic
inflationary models predict that these are mainly gaussian
adiabatic perturbations with a spectrum that is almost scale
invariant.

Inflationary dynamics  is typically studied by treating  the
inflaton as a homogeneous  classical scalar
field\cite{kolb,coles,liddle} whose evolution is determined
by a classical equation of motion,  while quantum fluctuations of
the scalar field around the classical value are treated in the
gaussian approximation and provide the seeds for the scalar
density perturbations of the metric. The quantum field theory
interpretation of the
 classical homogeneous field configuration that drives inflation is that
it is the expectation value
  of a quantum field operator in a translational invariant quantum
  state. There are important aspects of the dynamics that require a full
quantum treatment for their consistent description,
  for example particle production and in   particular
particle \emph{decay}.  A systematic treatment of the quantum
dynamics of the inflaton that includes particle production within
a non-perturbative framework is given in ref.\cite{cosmo,frw}.

While the dynamics of particle production during inflation has
received much attention, the full quantum treatment of particle
\emph{decay} during the inflationary (or more generally during a
rapidly expanding) stage has not been the focus of similar
attention.

While most reheating mechanisms rely on the coupling of the
inflaton to other fields into which it can decay leading to a
radiation dominated stage, the consequences of such coupling
between the inflaton and other fields \emph{during} the
inflationary stage are typically neglected. In this article we
focus on the study of the decay of a particle that could be the
inflaton as a result of its coupling to other fields.

To the best of our knowledge a preliminary study of the decay of
the inflaton during a de Sitter stage has been previously
addressed within a particular case in ref.\cite{prem}.

There could be several potentially important consequences of
particle decay during inflation: if the inflaton couples to other
particles, then its quantum fluctuations which seed density
perturbations also couple to these other fields. Therefore the
\emph{decay}  of  the quantum fluctuations of the inflaton may
result in a modification of the power spectrum of density
perturbations. Furthermore the coupling of the quantum
fluctuations of the inflaton and consequently of density
perturbations to other fields may possibly induce non-gaussian
correlations. For a recent review on non-gaussian correlations
generated during inflation see ref.\cite{bartolo}.

While a thorough  assessment of these potentially relevant
phenomena requires a detailed treatment of the coupling of gauge
invariant perturbations to other fields into which these
fluctuations can decay, in a spatially flat gauge there is a direct relation 
between the evolution equations for density perturbations and those 
of the quantum fluctuations of the inflaton\cite{hu2}.

Therefore if density perturbations also couple to these other
fields as a consequence of the coupling of the inflaton field to
these fields and \emph{if} density perturbations decay into these
 fields, such decay implies that the amplitude of density
perturbations will diminish with a  consequent modification of
their power spectrum.

Clearly a first step in the program to assess these potential
observables is to understand the decay of fluctuations
\emph{during} inflation which is the focus of our study in this
article.

These possibilities with distinct potential phenomenological
consequences for CMB anisotropies motivate us to study in detail
particle decay during an inflationary stage, which we take to be
described by a de Sitter space-time.

The decay of the inflaton during a post-inflationary stage  has
been considered recently\cite{7L} as a possible source of metric
(and therefore temperature) perturbations arising from an
inhomogeneity of the inflaton coupling. However most of these
treatments rely on  the concept of the decay rate of a particle
\emph{in Minkowski space-time } seemingly uncritical of  its
validity in the (rapidly) expanding universe.

\vspace{1mm}

{\bf Goals of this article:} In this article we study the decay of
a particle into other particles during inflation. The decaying
particle could be the inflaton but our study will be more
generally valid. The main focus of our study is to provide an
understanding of the concept of decay of a particle in a rapidly
expanding cosmology, and to introduce  and implement a method that
allows a systematic and unambiguous study of the relaxational
dynamics of quantum fields and in particular allows to extract the
decay law resulting from interactions.  In Minkowski space-time
there are two alternative but equivalent manners to define the
decay rate of a particle: I) the total decay rate is the inclusive
transition probability \emph{per unit time} from an initial 'in'
state to final 'out' states, II) the total decay rate is the
imaginary part of the space-time Fourier transform of the
self-energy of the particle evaluated on the particle's mass shell
and divided by its mass-shell energy. Both definitions are
equivalent by dint of the optical theorem, or alternatively,
unitarity. The calculation of a total decay rate from definition
I) involves calculating the transition amplitude from some initial
time $t_i \rightarrow -\infty$ to a final time $t_f \rightarrow
+\infty$ and multiplying by its complex conjugate. In Minkowski
space-time the transition amplitude from an asymptotic state in
the past to an asymptotic state in the future is proportional to
an energy conserving delta function. In squaring the amplitude,
the square of this delta function is interpreted as the total time
elapsed in the reaction ($T$) multiplying an energy conserving
delta function. Dividing by the total time of the reaction ($T$)
one extracts the decay rate. The calculation of the decay rate
from the total width via definition II) requires that the
self-energy be a function of the time difference and invokes
energy-momentum conservation at each interaction vertex. The
space-time Fourier transform of the self-energy features branch
cut singularities in the complex frequency plane and the imaginary
part across these cuts at the position of the particle mass shell
gives the decay width or decay rate. The important point in this
discussion is that in both cases the concept of a decay rate
relies heavily on energy (and momentum) conservation. Herein lies
the conceptual difficulty of extrapolating the concept of a decay
rate (an inclusive transition probability per unit time) to the
case of a rapidly expanding cosmology where there is no global
timelike Killing vector associated with conservation of energy
even when there may be space-like Killing vectors associated with
spatial translational symmetries and momentum conservation. Such
is the case for spatially flat Friedmann-Robertson-Walker
cosmologies. The manifest lack of energy conservation in an
expanding cosmology makes possible processes that would be
forbidden in a static space-time by energy conservation\cite{woodard}.
In addition, contrary to Minkowski spacetime, cosmological modes in general
do not decay exponentially with time, therefore the definition of the decay rate
requires the kind of analysis we provide here.

\vspace{1mm}

{\bf The method:}

Particle decay in de Sitter space-time has been previously studied
in reference\cite{prem} for some very special cases that allowed a
solution of the equation of motion. In this article we introduce a
method that allows to study the relaxation of quantum fields  and
particle decay  in great generality. The main strategy is to study
the effective equations of motion of the expectation values of
fields as an initial value problem in linear response including
the self-energy corrections. The solution of the equations of
motion lead to an unambiguous identification of the decay law from
the relaxation of the amplitude of the field as a consequence of
the self-energy corrections (interactions). When self-energy
corrections  are included the equations of motion become non-local
(non-Markovian) and cannot be solved in closed form in general.

When a perturbative solution of the equations of motion is
attempted there emerge \emph{secular terms}, namely terms that
grow in time and invalidate the perturbative expansion. These
secular terms indicate precisely the relaxation (or production)
time scales. We implement  the dynamical renormalization group
introduced in \cite{DRG} to provide a systematic resummation of
these secular terms leading to the correct description of
relaxation and decay. Such program has been successfully applied
to a wide variety of non-equilibrium situations in Minkowski space
time (see \cite{DRG} and references therein).

 In this article we generalize this approach to study particle
 decay during inflation. As a prelude to  studying more general situations,
we begin this program
 by implementing this
method to study the decay of a massive  and minimally coupled
particle into conformally coupled massless scalars via a trilinear
interaction vertex. After extracting the decay law to lowest order
in the loop expansion for the self-energy, we study the limit of
Minkowski space-time and show that the results obtained reproduce
those familiar in Minkowski space time.

{\bf Brief summary of results:} We introduce the dynamical
renormalization group\cite{DRG} method to study the relaxation of
the expectation value of quantum fields as an initial value
problem in the general case.

After introducing the  method and discussing its systematic
implementation in the general case we illustrate its application
and study the decay of a massive particle (it could be the
inflaton) coupled to
 conformally coupled massless particles via a trilinear vertex in de Sitter
space time.
 This simpler setting allows to present the main aspects of the
 program as well as reveal the important features associated with
 the expansion in a clear manner. The relaxation and decay law is
 studied to lowest order in the coupling both for wavelengths that
 are inside and outside the Hubble radius during inflation. The
 decay constant for superhorizon modes have an interesting
 interpretation in terms of the Hawking temperature of de
 Sitter space-time. In all cases we find that the decay
is \emph{enhanced} during inflation as compared to the Minkowsky
space-time result. The decay law for modes deep within the horizon
feature a wavevector dependence that leads to a larger suppression
of the amplitude for longer wavelengths.

The article is organized as follows: in section II  we introduce
the model and the non-equilibrium Green's and correlation
functions in arbitrary vacua, which are necessary ingredients for
obtaining the effective equations of motion. In section III we
obtain the equations of motion including self-energy corrections
in the loop expansion and introduce the dynamical renormalization
group method to extract the relaxation and decay law. In section
IV we study specific cases up to leading order in the
interaction and compare to the results in Minkowski space-time. In
section V we summarize our results and conclusions and
discuss potential implications of our results on
the power spectrum of density fluctuations
and non-gaussianity.

\section{The model}
We consider a spatially flat Friedmann-Robertson-Walker (FRW)
cosmological space time with scale factor $a(t)$, in comoving
coordinates the action is given by
\begin{equation}
A= \int d^3x \; dt \;  a^3(t) \Bigg\{
\frac{1}{2} \; {\dot{\phi}^2}-\frac{(\nabla
\phi)^2}{2a^2}-\frac{1}{2}\Big(M^2+\xi \; \mathcal{R}\Big)\phi^2 +
\frac{1}{2} \; {\dot{\varphi}^2}-\frac{(\nabla
\varphi)^2}{2a^2}-\frac{1}{2}\Big(m^2+\xi \; \mathcal{R}\Big)\varphi^2
- g \;  \phi \, \varphi^2 +J(t) \; \phi\Bigg\}
\end{equation}
\noindent with
\be \mathcal{R} = 6 \;  \Big(
\frac{\ddot{a}}{a}+\frac{\dot{a}^2}{a^2}\Big)
\ee
being the Ricci
scalar and $\xi$ an arbitrary coupling to the Ricci scalar: $\xi=
0$ corresponds to minimal coupling  and $\xi=1/6$ corresponds to
conformal coupling. The linear term in $\phi$ is a counterterm
that will be used to cancel the tadpole diagram in the equations
of motion.

It is convenient to pass to conformal time $\eta$ with $d\eta =
dt/a(t)$ and introduce a conformal rescaling of the fields
\begin{equation}
a(t) \; \phi(\vx,t) = \chi(\vx,\eta)~~;~~ a(t) \;
\varphi(\vx,t)=\delta(\vx,\eta)\;.
\end{equation}
The action becomes (after discarding surface terms that will not
change the equations of motion) \be A\Big[\chi,\delta\Big]=
\frac12 \int d^3x \; d\eta  \; \Bigg\{\frac12\left[
{\chi'}^2-(\nabla \chi)^2-\mathcal{M}^2_{\chi}(\eta) \; \chi^2 +
{\delta'}^2 -(\nabla \delta)^2-\mathcal{M}^2_{\delta}(\eta) \;
\delta^2 \right] -g C(\eta) \;  \chi \; \delta^2 - C^3(\eta) \;
J(\eta) \;  \chi \Bigg\} \; ,
\ee
\noindent with primes denoting derivatives with respect to
conformal time $\eta$ and \be \mathcal{M}^2_{\chi}(\eta) =
\Big(M^2+\xi_\chi \; \mathcal{R}\Big)
C^2(\eta)-\frac{C''(\eta)}{C(\eta)} \quad ,  \quad
 \mathcal{M}^2_{\delta}(\eta) =  \Big(m^2+\xi_\delta \; \mathcal{R}\Big)
C^2(\eta)-\frac{C''(\eta)}{C(\eta)}  \; ,
\ee
\noindent and $C(\eta)= a(t(\eta))$ is the scale factor as a
function of conformal time. For inflationary cosmology the scale
factor describes a de Sitter space-time, namely
\be
a(t)= e^{Ht} \; ,
\ee
\noindent with $H$ the Hubble constant and conformal time $\eta$
is given by
\be
\eta-\eta_0 = \frac{1}{H}(1-e^{-Ht})  \; ,
\ee
\noindent where $\eta_0$ corresponds to the initial time $t=0$. We
choose
\be
\eta_0 = -\frac{1}{H} \Rightarrow \eta= -\frac{e^{-Ht}}{H}~~;~~ C(\eta) =
-\frac{1}{H\eta}\label{scalefactor}  \; .
\ee
During inflation the effective time dependent masses of the fields are given by
\be
\mathcal{M}^2_{\chi}(\eta)  = \Big[\frac{M^2}{H^2}+12 \; \Big(\xi_\chi-
\frac{1}{6}\Big)\Big]\frac{1}{\eta^2}   \quad ,  \quad
 \mathcal{M}^2_{\delta}(\eta) =
\Big[\frac{m^2}{H^2}+12 \; \Big(\xi_\delta-\frac{1}{6}\Big)\Big]
\frac{1}{\eta^2} \; .
\ee
The Heisenberg equations of motion for the spatial Fourier modes
of wavevector $k$  of the fields in the non-interacting ($g=0$)
theory are given by
\bea \chi''_{\vk}(\eta)+
\Big[k^2-\frac{1}{\eta^2}\Big(\nu^2_{\chi}-\frac{1}{4} \Big)
\Big]\chi_{\vk}(\eta)& = & 0 \cr \cr
\delta''_{\vk}(\eta)+
\Big[k^2-\frac{1}{\eta^2}\Big(\nu^2_{\delta}-\frac{1}{4} \Big)
\Big]\delta_{\vk}(\eta)& = & 0 \label{deltamodes}  \; ,
\eea
\noindent where
\be
\nu^2_{\chi}  =  \frac{9}{4}-\Big( \frac{M^2}{H^2}+12 \; \xi_\chi
\Big) ~~;~~ \nu^2_{\delta}  = \frac{9}{4}-\Big(
\frac{m^2}{H^2}+12 \; \xi_\delta \Big) \label{nus}  \; .
\ee
The Heisenberg free field operators can be expanded in terms of
the linearly independent solutions of the mode equation
\be
S''_{\nu}(k;\eta)+ \Big[k^2-\frac{1}{\eta^2}\Big(\nu^2-\frac{1}{4} \Big)
\Big]S_{\nu}(k;\eta) =  0 \label{modeeqn} \; .
\ee
Two linearly independent solutions are given by
\bea
g_{\nu}(k;\eta) & = & \frac{1}{2}\; i^{-\nu-\frac{1}{2}}
\sqrt{\pi \eta}\,H^{(2)}_\nu(k\eta)\label{gnu}\\
f_{\nu}(k;\eta) & = & \frac{1}{2}\; i^{\nu+\frac{1}{2}}
\sqrt{\pi \eta}\,H^{(1)}_\nu(k\eta)= [g_{\nu}(k;\eta)]^*\label{fnu}  \; ,
\eea
 \noindent where $H^{(1,2)}_\nu(z)$ are Hankel functions. For
 wavevectors deep inside the Hubble radius $-k\eta >>1$
these functions have the asymptotic behavior
\be
g_{\nu}(k;\eta)  \buildrel{k\eta\rightarrow -\infty}\over=
\frac{1}{\sqrt{2k}} \; e^{-ik\eta} \quad , \quad
f_{\nu}(k;\eta)  \buildrel{k\eta\rightarrow -\infty}\over=
\frac{1}{\sqrt{2k}} \; e^{ ik\eta} \; ,
\label{fnuasy}
\ee
\noindent and are normalized so that their Wronskian is given by
\be
W[g_\nu(k;\eta),f_\nu(k;\eta)]=
g'_\nu(k;\eta)  \; f_\nu(k;\eta)-g_\nu(k;\eta) \; f'_\nu(k;\eta) = -i
\label{wronskian} \; .
\ee
The most general solution of the mode equations (\ref{modeeqn}) is
given by
\be
S_{\nu}(k;\eta)=
C_1(k;\eta_0) \; f_{\nu}(k;\eta)+C_2(k;\eta_0) \; g_{\nu}(k;\eta) \; ,
\label{gensol}
\ee
\noindent where the coefficients $C_{1,2}$ are determined by an
initial condition on the mode functions $S_{\nu}(k;\eta)$ at
conformal time $\eta_0$, namely
\bea
C_1(k;\eta_0)& = &
-i[g_{\nu}(k;\eta_0) \; S'_{\nu}(k;\eta_0)-g'_{\nu}(k;\eta_0)
\; S_{\nu}(k;\eta_0)] \label{C1} \\
C_2(k;\eta_0)& = &
-i[f'_{\nu}(k;\eta_0) \; S_{\nu}(k;\eta_0)-f_{\nu}(k;\eta_0) \;
S'_{\nu}(k;\eta_0)] \label{C2} \; .
\eea
The spatial Fourier transform of the free Heisenberg field
operators $ \chi_{\vk}(\eta), \; \delta_{\vk}(\eta) $ are therefore
written as
\bea \chi_{\vk}(\eta) & = &
\alpha_{\vk} \; S_{\nu_\chi}(k;\eta)+
\alpha^\dagger_{-\vk} \; S^*_{\nu_\chi}(k;\eta) \label{expchi}\\
\delta_{\vk}(\eta) & = &
\beta_{\vk} \; S_{\nu_\delta}(k;\eta)+\beta^\dagger_{-\vk} \;
S^*_{\nu_\delta}(k;\eta) \label{expdelta}  \; ,
\eea
\noindent where the Heisenberg operators
$\alpha_{\vk},  \; \alpha^\dagger_{\vk}$ and
$\beta_{\vk},  \; \beta^\dagger_{\vk}$ obey the usual canonical
commutation relations.

Canonical commutation relations both for the Heisenberg fields
(and their canonical momenta given by their derivatives with
respect to conformal time) and the creation and annihilation
operators entail that \be |C_2(k;\eta_0)|^2-|C_1(k;\eta_0)|^2 = 1
\label{constraint} \; . \ee The vacuum state $|0\rangle$ is
annihilated by $\alpha_{\vk},  \; \beta_{\vk}$. However a
different choice of the coefficients $C_{1,2}$ determine a
different  choice of the vacuum state, the Bunch-Davies vacuum
corresponds to choosing $C_2(k;\eta_0)=1, \; C_1(k;\eta_0)= 0
$ \cite{birr}. Heretofore
we generically refer to the different choices of the coefficients
$C_{1,2}$ constrained by Eq.  (\ref{constraint}),  as S-vacua.
 An illuminating representation of these coefficients can be
gleaned by computing the expectation value of the number operator
in the Bunch-Davies vacuum. Consider the expansion of a scalar field
either in terms of the Bunch-Davies basis $ g_k(\eta) $ or in terms
of the generalized basis $ S(k;\eta) $, namely
\begin{equation}\label{expa}
\chi_{\vk}(\eta)  = a_{\vk} \; g_{\nu_\chi}(k;\eta) +
a^{\dagger}_{-\vk}  \; g^*_{\nu_\chi}(k;\eta) =
\alpha_{\vk} \; S_{\nu_\chi}(k;\eta)+\alpha^\dagger_{-\vk} \;
S^*_{\nu_\chi}(k;\eta) \; ,
\end{equation}
\noindent the creation and annihilation operators are related by a
Bogoliubov transformation \be \alpha^\dagger_{\vk} =  C_2 \;
a^\dagger_{\vk} - C_1 \; a_{-\vk} \quad , \quad \alpha_{\vk} =
C^*_2 \; a_{\vk} - C^*_1 \; a^\dagger_{-\vk}  \; . \ee The
Bunch-Davies vacuum $|0\rangle_{BD}$ is annihilated by $a_{\vk}$,
hence we find the expectation value \be {}_{BD}\langle
0|\alpha^\dagger_{\vk}\alpha_{\vk}|0\rangle_{BD}= |C_1|^2= N_k \; .
\ee
Where $N_k$ is interpreted as the number of S-vacuum particles in
the Bunch-Davies vacuum. In combination with the constraint
(\ref{constraint}) the above result suggests the following
illuminating representation for the  coefficients $C_{1,2}$ \be
C_2(k) = \sqrt{1+N_k}~~;~~
 C_1(k)=\sqrt{N_k}\,e^{i\theta_k}
\ee
 \noindent where $N_k$ and $ \theta_k$  are real functions.

 \subsection{Non-equilibrium Green's and correlation functions}

 The main ingredients in the program to obtain the decay of the
 amplitude of the scalar field are the non-equilibrium Green's and
 correlation functions. Consider a generic free scalar field
 $\Phi$  quantized with the expansion
 \be \Phi_{\vk}(\eta) =
\alpha_{\vk} \; S_{\nu}(k;\eta)+\alpha^\dagger_{-\vk} \; S^*_{\nu}(k;\eta) \; .
\ee
The non-equilibrium Green's and correlation functions are given by
\bea
G^{++}_k(\eta,\eta') & = & \langle 0|
T\left(\Phi_{\vk}(\eta)\Phi_{-\vk}(\eta') \right) |0\rangle =
S_{\nu}(k;\eta) \; S^*_{\nu}(k;\eta') \; \Theta(\eta-\eta')+
S^*_{\nu}(k;\eta) \; S_{\nu}(k;\eta') \; \Theta(\eta'-\eta) \label{G++}\\
G^{--}_k(\eta,\eta') & = & \langle
0|\widetilde{T}\left(\Phi_{\vk}(\eta)\Phi_{-\vk}(\eta')\right)
|0\rangle = S_{\nu}(k;\eta) \; S^*_{\nu}(k;\eta') \; \Theta(\eta'-\eta)+
S^*_{\nu}(k;\eta) \; S_{\nu}(k;\eta') \; \Theta(\eta-\eta') \label{G--}\\
G^{-+}_k(\eta,\eta') & = & \langle 0|
\Phi_{\vk}(\eta)\Phi_{-\vk}(\eta') |0\rangle =
S_{\nu}(k;\eta) \; S^*_{\nu}(k;\eta')\label{G-+}
\\
G^{+-}_k(\eta,\eta') & = & \langle 0|
\Phi_{-\vk}(\eta')\Phi_{\vk}(\eta) |0\rangle = \left[
G^{-+}_k(\eta,\eta')\right]^* =
S^*_{\nu}(k;\eta) \; S_{\nu}(k;\eta')\label{G+-}  \; ,
\eea
\noindent where $T,\widetilde{T}$ stand for the time and anti-time
ordering symbols respectively. These Green's and correlation
functions are \emph{not} independent since they fulfill the
following identity
\be\label{identit}
G^{++}_k(\eta,\eta') + G^{--}_k(\eta,\eta')=
G^{-+}_k(\eta,\eta')+G^{+-}_k(\eta,\eta')   \; ,
\ee \noindent which can be trivially verified.

For generalized S-vacua, we find for example
\be
G^{+-}_k(\eta,\eta')=
[1+N_k] \; g^*_\nu(k;\eta) \; g_\nu(k;\eta')+N_k \; g_\nu(k;\eta) \;
g^*_\nu(k;\eta')+\sqrt{[1+N_k]N_k} \; \left[e^{-i\theta_k}g_\nu(k;\eta) \;
g_\nu(k;\eta')+~c.c.\right]  \; .
\ee
For physical wavelengths
that are much smaller than the Hubble radius, namely for
$k\eta,k\eta' \gg 1$ the two point correlation function above has
the following behavior
\be G^{+-}_k(\eta,\eta') \buildrel{k\eta,k\eta'\rightarrow
\infty}\over= \frac{1+N_k}{2k} \;
e^{-ik(\eta-\eta')}+\frac{N_k}{2k} \;
e^{ik(\eta-\eta')}+\frac{1}{k} \; \sqrt{[1+N_k]N_k} \;
\cos\left[k(\eta+\eta')+\theta_k\right)] \; .
\ee
The first two terms are similar to the Wightman function of a
scalar field  in a bath  in equilibrium, whereas the last terms
with the \emph{sum} of the conformal times are a distinct feature
of the mixing between particle and antiparticle states of the
Bunch-Davies vacuum. We can impose asymptotically, for
wavelengths deep within the Hubble radius that the physics be locally that of
flat Minkowski space-time with a timelike Killing vector which implies
that the two point function be translational invariant in time.
This condition  requires that the occupation numbers fulfill $N_k
\rightarrow 0$ as $k\rightarrow \infty$. In particular if we
further demand that the number of Bunch-Davies particles in a
generalized S-vacuum be finite, then it must be that $N_k \lesssim
{\cal O} \left( 1/k^{3+\epsilon}\right), \; \epsilon>0 $ as
$k\rightarrow \infty$.

\section{Equations of motion and the dynamical renormalization group}

As mentioned in the introduction, in Minkowski space-time the
decay rate of a particle can be obtained either from the imaginary
part of the space-time Fourier transform of the retarded
self-energy or alternatively from the transition probability per
unit time. These methods are equivalent by dint of the optical
theorem (unitarity) and both rely on energy conservation. In a
rapidly expanding cosmology energy is not conserved (albeit there
is covariant conservation) and the self-energy is not a function
of the time difference but is a rapidly varying non-local function
of the time arguments. In this situation  a decay `rate' cannot
be identified unambiguously.

In ref.\cite{DRG} we have introduced a method to study relaxation
directly in \emph{real time}. The method relies in obtaining
the equation of motion of expectation values of the fields in
linear response. The decay law (more general than a decay
rate) is extracted from the time evolution of the
\emph{amplitude} of the expectation value. The equation of motion
for the expectation value includes the non-local self-energy
contributions in a consistent loop expansion. Although the
equation is linear, it is generally non-local and when the
self-energy is not time translational invariant (no energy
conservation) it becomes an integro-differential equation which in
general cannot be solved in closed form. The method introduced in
ref.\cite{DRG} relies on a \emph{perturbative} expansion of the
solution in terms of the coupling constant. However, such an
expansion features secular terms, namely terms that grow in time
and invalidate the perturbative expansion. The dynamical
renormalization group\cite{DRG} provides a consistent resummation
of the series that leads to a uniform asymptotic expansion. The
DRG improved solution directly allows to extract the decay law.
This method has been applied and its applicability and reliability
has been tested  in a variety of equilibrium and non-equilibrium
situations. The method is rather general and allows to resum 
secular terms for  \emph{any} set of differential or
integro-differential evolution equations.

While the method has not yet been applied to the case of an
expanding cosmology, we will confirm its reliability by analyzing
the results in the limit when the expansion rate vanishes, namely,
Minkowski space-time.

The method to study decay and relaxation begins by obtaining the
equations of motion for the expectation value of the decaying
field, which is obtained from the  non-equilibrium generating
functional\cite{cosmo,frw}
\be
\mathcal{Z} = \int
\mathcal{D\chi^{\pm}} \; \mathcal{D\delta^{\pm}} \;  e^{i\left\{
A[\chi^+,\delta^+]-A[\chi^-,\delta^-]\right\}}  \; .
\ee
Where the superscripts $\pm$ refer to the forward ($+$) and
backward ($-$) branches along a closed contour in time
corresponding to the unitary time evolution operator and its
inverse, both of which are required to obtain expectation values
in an initial density matrix (which could describe a pure state).
For details on this method the reader is referred to
ref.\cite{cosmo,frw}.

Our goal is to obtain the equation of motion for the expectation
value of the field $\chi$. For this purpose, we implement the
tadpole method by performing the following shift in the spatial
Fourier transform of the field $\chi$\cite{cosmo,frw} \be
\label{shift} \chi^{\pm}_{\vk}(\eta) = X_{\vk}(\eta)+
\sigma^{\pm}_{\vk}(\eta)~~;~~\langle \chi^{\pm}_{\vk}(\eta)\rangle
= X_{\vk}(\eta)~~;~~\langle  \sigma^{\pm}_{\vk}(\eta)\rangle =0 \; .
\ee
\noindent in the above expressions $\langle \cdots \rangle$ stand
for expectation values in the initial state which can be prepared
by switching on an external source term to displace the field and
switching the source off to let the field evolve. This is the
usual method to prepare an initial value problem in linear
response. For more details on this method we refer the reader
to\cite{cosmo,DRG}.

\begin{figure}
\includegraphics[height=3 in,width=3 in,keepaspectratio=true]{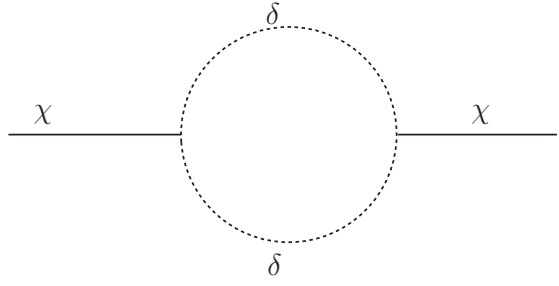}
\caption{Self energy loop of conformally coupled massless
particles (dashed lines). } \label{fig:selfener}
\end{figure}

In terms of the shifted field, the action for the spatial Fourier
transformed fields becomes \bea &&A[X,\sigma^\pm,\delta^\pm] =
\frac12\int_{\eta_0}^0 d\eta \sum_{\vk} \Bigg\{
\sigma^{'\pm}_{\vk} \; \sigma^{'\pm}_{-\vk}-
\left[k^2+\mathcal{M}^2_{\chi}(\eta)\right]
\sigma^{\pm}_{\vk}\sigma^{\pm}_{-\vk}+ \delta^{'\pm}_{\vk} \;
\delta^{'\pm}_{-\vk}-\left[k^2+
\mathcal{M}^2_{\delta}(\eta)\right]\delta^{\pm}_{\vk} \;
\delta^{\pm}_{-\vk}
\Bigg\} \nonumber\\
&  & -\int_{\eta_0}^0 d\eta  \; \sum_{\vk} \Bigg\{
\sigma^{\pm}_{-\vk}\Big[X''_{\vk}+\left(k^2+
\mathcal{M}^2_{\chi}(\eta)\right)X_{\vk}\Big]+ g~C(\eta) \;
\frac{1}{\sqrt{V}} \;
\sum_{\vec{q}}\Big(\sigma^{\pm}_{\vk}+X_{\vk}\Big) \;
\delta_{\vec{q}} \; \delta_{-\vk-\vec{q}}-C^3(\eta) \; J(\eta) \;
X_{\vec{0}}\Bigg\}\nonumber \eea \noindent where we have neglected
terms that cancel in the difference
$A[X,\sigma^+,\delta^+]-A[X,\sigma^-,\delta^-]$, and $V$ is the
comoving spatial volume. The equation of motion for the
expectation value $X_{\vk}(\eta)$ is obtained by implementing the
condition $\langle\sigma^{\pm}(\eta)\rangle=0$ order by order in
perturbation theory. The equation of motion is the same for the
$+$ and $-$ conditions as a consequence of the identity
(\ref{identit}), and up to one loop order it is given by \be
\label{eqnofmot}
X''_{\vk}(\eta)+\left[k^2+\mathcal{M}^2_{\chi}(\eta)\right]
X_{\vk}(\eta)+ 2 \, g^2 \; C(\eta) \; \int_{\eta_0}^{\eta}
d\eta'~C(\eta') \; \mathcal{K}_k(\eta,\eta') \;
X_{\vk}(\eta')=0\;, \ee
\noindent where we have used the
counterterm $J(\eta)$ to cancel the tadpole $\langle
\delta_{\vec{q}}\,\delta_{-\vec{q}}\rangle$. At  one loop order in
the fields $\delta$ ($\mathcal{O}(g^2)$) the non-local kernel
$\mathcal{K}_k(\eta,\eta')$ is given  by the following expression
\be\label{kernel} \mathcal{K}_k(\eta,\eta')  = -i \int
\frac{d^3q}{(2\pi)^3} \Big[G^{-+}_{\delta,\vec{q}}(\eta,\eta') \;
G^{-+}_{\delta,\vec{q}+\vk}(\eta,\eta')-G^{+-}_{\delta,\vec{q}}(\eta,\eta')
 \; G^{+-}_{\delta,\vec{q}+\vk}(\eta,\eta')\Big]\;,
\ee \noindent where the  correlation functions
$G^{\pm,\mp}_{\delta,\vec{q}}(\eta,\eta')$ are those for the
$\delta$ field given by the expressions
Eqs.(\ref{G++})-(\ref{G+-}) with $\nu=\nu_{\delta}$ given in Eq.
(\ref{nus}). This non-local kernel corresponds to the one-loop
retarded self energy in real time as shown in fig. \ref{fig:selfener}.

Although the equation of motion (\ref{eqnofmot}) is linear, it is
non-local and a general solution is unavailable. However for weak
coupling $g^2$ a perturbative solution can be found by writing
\be
X_{\vk}(\eta)= \sum_{n=0}^{\infty}(g^2)^n~X_{n,\vk}(\eta) \; ,
\ee
\noindent leading to the hierarchy of coupled equations
\begin{eqnarray}\label{perteqn}
&&X''_{0,\vk}(\eta)+\left[k^2-\frac{1}{\eta^2}\Big(\nu^2_{\chi}-
\frac{1}{4} \Big) \right]X_{0,\vk}(\eta)
 =  0  \; ,\\
&&X''_{n,\vk}(\eta)+\left[k^2-\frac{1}{\eta^2}\Big(\nu^2_{\chi}-
\frac{1}{4} \Big) \right]
X_{n,\vk}(\eta) = \mathcal{R}_n(k;\eta) ~~;~~n=1,2\cdots\\
&&  \mathcal{R}_n(k;\eta) =   - 2 \; C(\eta)\int_{\eta_0}^{\eta}
d\eta'~C(\eta') \; \mathcal{K}_k(\eta,\eta') \; X_{n-1,\vk}(\eta') \; .
\label{rn}
\end{eqnarray}
In terms of two linearly independent solutions of the unperturbed
equation $g_{\nu_\chi}(k,\eta);f_{\nu_\chi}(k,\eta)$ with
Wronskian [see Eq. (\ref{wronskian})] equal to $ - i $, we write
\be\label{solzero} X_{0,\vk}(\eta)= A_{\vk} \;
g_{\nu_\chi}(k;\eta)+B_{\vk} \; f_{\nu_\chi}(k;\eta) \; , \ee
\noindent where the coefficients $A_{\vk}, \; B_{\vk}$ are
determined by initial conditions. The hierarchy of coupled
equations (\ref{perteqn}) can be solved iteratively by introducing
the (retarded) Green's function of the second order differential
operator on the left hand side of these equations. This Green's
function is given by \be\label{GF} \mathcal{G}(k;\eta,\eta')=i
\left[g_{\nu_\chi}(k;\eta) \; f_{\nu_\chi}(k;\eta')-
f_{\nu_\chi}(k;\eta) \; g_{\nu_\chi}(k;\eta')
\right]\Theta(\eta-\eta')\;. \ee Hence the solution of the
hierarchy of equations for $n\geq 1$ is given by \be
\label{forsol} X_{n,\vk}(\eta)= \int_{\eta_0}^{0} d\eta' \;
\mathcal{G}(k;\eta,\eta') \; \mathcal{R}_{n}(k;\eta') \; . \ee The
general form of the perturbative solution Eq.(\ref{forsol})
combined with the linearity of the equation of motion indicate
that the full solution is of the form \be X_{\vk}(\eta)= A_{\vk}
\; g_{\nu_\chi}(k;\eta)\left[1+ g^2 \; F_1 (k;\eta)+g^4 \;
F_2(k;\eta)+{\cal O}(g^6) \right]+B_{\vk} \;
f_{\nu_\chi}(k;\eta)\left[1+ g^2 \;  H_1 (k;\eta)+g^4 \;
H_2(k;\eta)+{\cal O}(g^6) \right] \label{solugen} \ee \noindent
where the functions $F_{n}(k,\eta); \; H_{n}(k;\eta)$ are found
iteratively from the procedure described above. If the functions
$F_{n}(k;\eta); \; H_{n}(k;\eta)$ remain bounded in the limit
$\eta \rightarrow 0$ then the perturbative expansion provides a
convergent uniform series. However, in general these functions
feature \emph{secular terms}, namely contributions that
\emph{diverge} in the limit $\eta\rightarrow 0$ and invalidate the
perturbative expansion. In order to provide a uniform solution
valid at all (conformal) times, these secular terms must be
resummed via the dynamical renormalization group\cite{DRG}.

\subsection{Dynamical renormalization group (DRG)}\label{DRG}

The general solution given by Eq. (\ref{solugen}) highlights that
the perturbative corrections can be interpreted as a
\emph{renormalization} of the complex amplitudes
$A_{\vk};B_{\vk}$, the dynamical renormalization group provides a
systematic manner to resum the secular divergences in the
perturbative expansion in terms of renormalization of the
amplitudes.

This resummation program begins by extracting the secular terms of
the functions $F_{n}(k;\eta), \; H_{n}(k;\eta)$ from the terms
that remain finite and bounded in conformal time. Let us write \be
F_{n}(k;\eta)= F_{n,s}(k;\eta)+F_{n,f}(k;\eta)~~;~~ H_{n}(k;\eta)=
H_{n,s}(k;\eta)+H_{n,f}(k;\eta)\;,\label{extract} \ee \noindent
where $F_{n,s}(k;\eta), \; H_{n,s}(k;\eta)$ are secular, namely
diverge in the limit $\eta \rightarrow 0$, whereas
$F_{n,f}(k;\eta), \; H_{n,f}(k;\eta)$ remain bounded for  $\eta
\rightarrow 0$.

The dynamical renormalization group implements a resummation of
the secular terms by introducing an arbitrary scale $\tilde{\eta}$
and a wave function renormalization of the complex amplitudes as
follows\cite{DRG}
 \bea A_{\vk} & = &  A_{\vk}(\tilde{\eta}) \;
Z^A_{\vk}(\tilde{\eta}) ~;~ Z^A_{\vk}(\tilde{\eta})= 1+ g^2 \;
z^A_{1,\vk}(\tilde{\eta})+ g^4 \;
z^A_{2,\vk}(\tilde{\eta})+\mathcal{O}(g^6)\;, \label{ZA}\\
B_{\vk} & = &  B_{\vk}(\tilde{\eta} \; )Z^B_{\vk}(\tilde{\eta})
~;~ Z^B_{\vk}(\tilde{\eta})= 1+ g^2 \;  z^B_{1,\vk}(\tilde{\eta})+
g^4 \; z^B_{2,\vk}(\tilde{\eta})+\mathcal{O}(g^6) \;.\label{ZB}
\eea The solution of the equation of motion now becomes \be
X_{\vk}(\eta)   = A_{\vk}(\tilde{\eta}) \;
g_{\nu_\chi}(k;\eta)\left[1+ g^2 \left(F_1
(k;\eta)+z^A_{1,\vk}(\tilde{\eta})\right)+\mathcal{O}(g^4)
\right]+ B_{\vk}(\tilde{\eta}) \; f_{\nu_\chi}(k;\eta)\left[1+
g^2\left(H_1
(k;\eta)+z^B_{1,\vk}(\tilde{\eta})\right)+\mathcal{O}(g^4) \right]
\label{solugenren} \ee The coefficients $z^{A,B}_{n,\vk}$ are
chosen so that they cancel the secular terms $F_{n,s}(k;\eta), \;
F_{n,s}(k;\eta)$ at the point  $\eta = \tilde{\eta}$, namely \be
z^A_{1,\vk}(\tilde{\eta})= -F_{1,s}(k;\tilde{\eta})~~;~~
z^B_{1,\vk}(\tilde{\eta})= -H_{1,s}(k;\tilde{\eta})~~;~~
\mathrm{etc}. \ee Therefore, the solution now becomes \be
X_{\vk}(\eta)   = A_{\vk}(\tilde{\eta}) \;
g_{\nu_\chi}(k;\eta)\left\{ 1+ g^2 \left[F_1 (k;\eta)-F_{1,s}
(k;\tilde{\eta})\right]+\mathcal{O}(g^4) \right\}+
B_{\vk}(\tilde{\eta}) \; f_{\nu_\chi}(k;\eta)\left\{ 1+
g^2\left[H_1 (k;\eta)-H_1
(k;\tilde{\eta})\right]+\mathcal{O}(g^4)\right\} \ee This form of
the solution can be written in the more illuminating manner
\bea\label{solimpro} X_{\vk}(\eta)   = && A_{\vk}(\tilde{\eta}) \;
g_{\nu_\chi}(k;\eta)\left\{ 1+ g^2 \int^{\eta}_{\tilde{\eta}}
\frac{dF_{1,s} (k;\eta')}{d \eta'}
\,d\eta'+\mathcal{O}(g^4)+\mbox{non-secular} \right\}+ \nonumber\\
&+& B_{\vk}(\tilde{\eta}) \; f_{\nu_\chi}(k;\eta)\left\{ 1+ g^2
\int^{\eta}_{\tilde{\eta}} \frac{dH_{1,s} (k;\eta')}{d
\eta'}\,d\eta'+\mathcal{O}(g^4)+\mbox{non-secular}\right\} \eea
\noindent where the non-secular terms are terms {\bf bounded} in
the limit $\eta \rightarrow 0$. Eq.(\ref{solimpro}) reveals that
the solution has been improved by choosing the scale
$\tilde{\eta}$ arbitrarily close to $\eta$.

 The solution $X_{\vk}(\eta)$ is
\emph{independent} of the arbitrary renormalization scale
$\tilde{\eta}$, namely \be \frac{d X_{\vk}(\eta)}{d\tilde{\eta}}=0
\ee \noindent which leads to the \emph{dynamical renormalization
group equation}\cite{DRG}.  To lowest order  the DRG equation is
given by

\bea \frac{\partial A_{\vk}(\tilde{\eta})}{\partial
\tilde{\eta}}-g^2 \; A_{\vk}(\tilde{\eta}) \; \frac{\partial
F_{1,s} (k;\tilde{\eta})}{\partial \tilde{\eta}}+\mathcal{O}(g^4)
& = & 0 \;, \label{DRGA}\\
\frac{\partial B_{\vk}(\tilde{\eta})}{\partial \tilde{\eta}}-g^2
\; B_{\vk}(\tilde{\eta}) \; \frac{\partial H_{1,s}
(k;\tilde{\eta})}{\partial \tilde{\eta}}+\mathcal{O}(g^4)  & = & 0
 \;, \label{DRGB}\eea

To lowest order in $g^2$ the solution of these dynamical
renormalization group equations is given by \bea
A_{\vk}(\tilde{\eta})= A_{\vk}(\tilde{\eta}_0)~e^{g^2\left[F_{1,s}
(k;\tilde{\eta})-F_{1,s} (k;\tilde{\eta}_0) \right]}\\
B_{\vk}(\tilde{\eta})= B_{\vk}(\tilde{\eta}_0)~e^{g^2\left[H_{1,s}
(k;\tilde{\eta})-H_{1,s} (k;\tilde{\eta}_0) \right]}\;. \eea Since
the scale $\tilde{\eta}$ is arbitrary, we can now set
$\tilde{\eta}=\eta$ and obtain the renormalization group improved
solution \bea X_{\vk}(\eta) & = & A_{\vk}({\eta})
\,g_{\nu_\chi}(k;\eta)\left[1+g^2
F_{1,f}(k,\eta)+\mathcal{O}(g^4)\right]+B_{\vk}({\eta}) \;
f_{\nu_\chi}(k;\eta)\left[1+g^2
H_{1,f}(k,\eta)+\mathcal{O}(g^4)\right]\;,\label{RGsol}\\
A_{\vk}({\eta})& = &  A_{\vk}({\eta}_0)~e^{g^2\left[F_{1,s}
(k;{\eta})-F_{1,s} (k;{\eta}_0) \right]}\;,\label{Aamp}\\
 B_{\vk}({\eta}) & = &
B_{\vk}({\eta}_0)~e^{g^2\left[H_{1,s} (k;{\eta})-H_{1,s}
(k;{\eta}_0) \right]}\;,\label{Bamp} \eea \noindent the terms in
the brackets in Eq. (\ref{RGsol}) are truly perturbatively small
at all conformal times for weak coupling. That is, the dynamical
renormalization group produces a perturbative expansion which is
uniform in time. The exponential factors in the complex amplitudes
(\ref{Aamp})-(\ref{Bamp}) will determine the decay of the
amplitude. The reliability and power of this method have been
tested in many different cases and we refer the reader
to\cite{DRG} for a more thorough discussion.

We now implement this program in several relevant cases.

\section{Specific cases}

We begin our program by implementing the dynamical renormalization
group resummation in a simpler case and to lowest order in the
coupling, namely $\mathcal{O}(g^2)$, with the goal of highlighting
the main aspects of the program in a simpler setting.

For this we consider the $\delta$ field (decay product) to be a
massless conformally coupled field in its Bunch-Davies vacuum,
namely $m=0, \; \xi_\delta =1/6 , \; C_2=1$ and the decaying field
$\chi$ to be massive and minimally coupled, namely $\xi_\chi =0$ .
For the case of massless conformally coupled particles with
Bunch-Davies vacuum in the self-energy loop \be
S_{\nu_{\delta}}(k;\eta)= \frac{1}{\sqrt{2k}} \;
e^{-ik\eta}\label{confmode} \ee \noindent and the non-local kernel
is given by \be\label{kernelconf} \mathcal{K}_k(\eta,\eta') =
\int\frac{d^3q}{(2\pi)^3}
\frac{\sin\left[(q+|\vec{k}+\vec{q}|)(\eta-\eta')\right]}{2q|\vec{k}+\vec{q}|}
=-\frac{1}{8\pi^2}\cos[k(\eta-\eta')]~\mathcal{P}
\left(\frac{1}{\eta-\eta'} \right) \ee \noindent where
$\mathcal{P}$ stands for the principal part. We \emph{define} the
principal part prescription as follows \be\label{PP} \mathcal{P}
\left(\frac{1}{\eta-\eta'} \right) \equiv
\frac{\eta-\eta'}{\left(\eta-\eta'\right)^2 + (\epsilon \;
\eta')^2}= \frac12\left[\frac{1}{\eta-\eta'+i\epsilon \; \eta'}+
\frac{1}{\eta-\eta'-i\epsilon \; \eta'}\right] ~~;~~ \epsilon
\rightarrow 0 \ee This prescription for the principal part
regulates the short distance divergence in the operator product
expansion with a dimensionless infinitesimal quantity $\epsilon$
independent of time. This $\eta'$-dependent point-splitting
prescription in conformal time correspond to a time-independent
point-splitting in comoving time $ t = -\frac{1}{H} \log[-H\;
\eta] $ [see Eq.(\ref{scalefactor})]: \be
\frac{\eta-\eta'}{\left(\eta-\eta'\right)^2 + (\epsilon \;
\eta')^2}= H\; e^{H\;t'} \; \frac{1 - e^{-H(t-t')}}{\left[1 -
e^{-H(t-t')}\right]^2+ \epsilon^2}. \label{pointsp} \ee

That is, a time splitting of $ \epsilon/H $  between the points
$t$ and $t'$ for $ t \to t' $. This choice of regularization is
consistent with the short-distance singularities of the operator
product expansion in Minkowski space-time, and  leads to a
time-independent mass renormalization. Indeed, time-dependent mass
renormalizations are allowed in cosmological space-times and they
are associated with different regularization prescriptions.
(For an analogous discussion using the moment cutoff instead of
point splitting see sec. III of ref. \cite{frw}).

Repeating the calculations that follow but with  a (conformal)
time independent point-splitting $\epsilon$ instead of $\epsilon
\; \eta'$ as conformal time separation in Eq.(\ref{PP}) leads to a
(conformal) time dependent mass renormalization. While there is no
unique choice of renormalization prescription, we demand a time
independent renormalization of the mass, which is achieved by the
principal part prescription adopted in Eq. (\ref{PP}).

Even with the simplification of conformally coupled massless
fields in the self-energy loop, the study of the general case for
arbitrary wavevectors is complicated by the fact that the
solutions of the equations of motion at zero order are given by
Eq. (\ref{solzero}) with $ g_{\nu_{\chi}}(k;\eta) $ given by Eq.
(\ref{gnu}) and $ \nu_{\chi} $ given by Eq. (\ref{nus}) for $
\xi_{\chi}=0 $. However, progress can be made in the following
relevant cases: i) $ k\eta,k\eta' \ll 1 $ corresponding to
wavelengths that are larger than the Hubble radius all throughout
inflation, which is equivalent to taking $ k=0 $, ii) $k\neq 0$
with $|k\eta| \rightarrow 0$, this corresponds to  modes that
cross the horizon during inflation and iii) $ k\eta, \; k\eta' \gg
1 $ corresponding to wavelengths that are smaller than the Hubble
radius all throughout inflation. We study each case separately.

\subsection{Wavelengths larger than the Hubble radius: $k=0$}

In the case $k=0$ it is convenient to choose the following
linearly independent solutions of the unperturbed equations of motion
\be
g_{\nu_{\chi}}(0,\eta)   =  (-\eta)^{\beta_+} \quad ,
\quad f_{\nu_{\chi}}(0,\eta)   =  (-\eta)^{\beta_-} \; , \ee
\noindent with \be \beta_{\pm}  =   \frac{1}{2}\pm
\nu_{\chi}~~,~~\nu_{\chi}=
\sqrt{\frac{9}{4}-\frac{M^2}{H^2}}\label{betapm}  \; ,
\ee
\noindent in terms of which we write the solutions for the
equation of motion at zeroth order as
\be\label{X0}
X_{0,\vec{0}}(\eta) = a \; (-\eta)^{\beta_+}+ b \;
(-\eta)^{\beta_-} \; ,
\ee
where $ a $ and $ b $ are constant coefficients.

The retarded Green's function
Eq.(\ref{GF}) necessary to solve the hierarchy of coupled
equations, for $k=0$ is given by \be\label{GFk0}
\mathcal{G}(\eta,\eta')=
-\frac{1}{2\nu_{\chi}}\left[(-\eta)^{\beta_+}~(-\eta')^{\beta_-}-
(-\eta')^{\beta_+}~(-\eta)^{\beta_-} \right]\Theta(\eta-\eta')=
\frac{\sqrt{\eta \; \eta'}}{2 \; \nu_{\chi}}
\left[\left(\frac{\eta'}{\eta}\right)^{\nu_{\chi}}-
\left(\frac{\eta}{\eta'}\right)^{\nu_{\chi}}\right]\Theta(\eta-\eta')
\; .
\ee
We start by computing $ \mathcal{R}_1(\vec{0},\eta) $ which from
eqs. (\ref{rn}), (\ref{PP}) and (\ref{X0}) is given by
$$
\mathcal{R}_1(\vec{0},\eta)= -\frac{1}{8 \; \pi^2 \; H^2 \; \eta}
\int_{\eta_0}^{\eta} \frac{d\eta'}{\eta'}
\frac{a \; (-\eta')^{\beta_+}+ b \; (-\eta')^{\beta_-}}{\eta-\eta'+
i\epsilon \; \eta'} + (\epsilon \to - \epsilon) \; .
$$
Expanding the kernel in a power series of the ratio $\eta/\eta'$ and
integrating term by term yields
\bea\label{desR1}
&&\mathcal{R}_1(\vec{0},\eta) = -\frac{1}{(2 \; \pi \; H \; \eta)^2}
\left\{ \left[ a \; (-\eta)^{\beta_+}+ b \; (-\eta)^{\beta_-} \right]
(\log\epsilon + \gamma ) + a \left[  (-\eta)^{\beta_+} \; \psi(1-\beta_+)
+  (-\eta_0)^{\beta_+} \; \sum_{k=1}^{\infty} \frac{1}{k -\beta_+}
\left(\frac{\eta}{\eta_0} \right)^k  \right]+\right. \cr \cr
&& \left. + b \left[  (-\eta)^{\beta_-} \; \psi(1-\beta_-)
+  (-\eta_0)^{\beta_-} \; \sum_{k=1}^{\infty} \frac{1}{k -\beta_-}
\left(\frac{\eta}{\eta_0} \right)^k  \right] \right\} \; ,
\eea
where $ \gamma =  0.57721\ldots $ is the Euler-Mascheroni constant and
$ \psi(z) $ stands for the logarithmic derivative of the Gamma function.

Inserting Eqs.(\ref{GFk0}) and (\ref{desR1}) in Eq. (\ref{forsol})
we find that the first order correction $X_{1,\vec{0}}(\eta)$ is
given by \bea &&X_{1,\vec{0}}(\eta)=- \frac{1}{2 \; (2 \; \pi \;
H)^2 \; \nu_{\chi}} \left( a \; (-\eta)^{\beta_+} \left\{ \left[
\log\epsilon + \gamma +  \psi(1-\beta_+) \right] \log
\frac{\eta}{\eta_0} + \mbox{non-secular}\right\} \right. \cr \cr
&&\left. - b \; (-\eta)^{\beta_-}  \left\{ \left[ \log\epsilon +
\gamma +  \psi(1-\beta_-) \right] \log \frac{\eta}{\eta_0} +
\mbox{non-secular}\right\} \right)\;, \eea
\noindent where the
non-secular terms are terms {\bf bounded} in the limit $\eta
\rightarrow 0$. Therefore, to first  order in the coupling we find
the solution of the equation of motion for the $\vk=\vec{0}$ mode
to be given by \bea\label{solX0} &&X_{\vec{0}}(\eta) = a \;
(-\eta)^{\beta_+}\left(1-\frac{g^2}{2 \; (2 \; \pi \; H)^2 \;
\nu_{\chi}} \left\{ \left[ \log\epsilon + \gamma +  \psi(\beta_-)
\right] \log \frac{\eta}{\eta_0} + \mbox{non-secular}\right\}
\right) +\cr \cr &&+b \; (-\eta)^{\beta_-}\left(1+\frac{g^2}{2 \;
(2 \; \pi \; H)^2 \; \nu_{\chi}} \left\{ \left[ \log\epsilon +
\gamma +  \psi(\beta_+) \right] \log \frac{\eta}{\eta_0} +
\mbox{non-secular}\right\} \right) \; . \eea where we used that $
\beta_+ + \beta_- = 1 $ [Eq.(\ref{betapm})]. From this expression
we can read off the secular contributions $F_{1,s}(0,\eta), \;
H_{1,s}(0,\eta)$ in Eq. (\ref{extract}) for $k=0$,  namely \bea
F_{1,s}(0,\eta) & = &-\frac{1}{2 \; (2 \; \pi \; H)^2 \;
\nu_{\chi}}
 \left[\log\epsilon + \gamma +  \psi(\beta_-) \right]\ln\frac{\eta}{\eta_0}
\label{k0secf}\\
H_{1,s}(0,\eta) & = &
 \frac{1}{2 \; (2 \; \pi \; H)^2 \; \nu_{\chi} }
 \left[\log\epsilon + \gamma +  \psi(\beta_+) \right]\ln\frac{\eta}{\eta_0}
\label{k0sech} \eea And the dynamical renormalization group
resummation analyzed in section (\ref{DRG}) above leads to the
following resummed  solution \be X_{\vec{0}}(\eta)=
a(\tilde{\eta}_0) \; (-\eta)^{\beta_+}
\left[\frac{\eta}{\tilde{\eta}_0}\right]^{-\frac{g^2\left[\log\epsilon
+ \gamma + \psi(\beta_-) \right]}{2 \; (2 \; \pi \; H)^2 \;
\nu_{\chi}}} [1+\mathcal{O}(g^2)]+b(\tilde{\eta}_0) \;
(-\eta)^{\beta_-}
\left[\frac{\eta}{\tilde{\eta}_0}\right]^{\frac{g^2\left[\log\epsilon
+ \gamma + \psi(\beta_+) \right]}{2 \; (2 \; \pi \; H)^2 \;
\nu_{\chi}}} [1+\mathcal{O}(g^2)] \;,\label{resumX0} \ee \noindent
where the terms in the brackets are perturbatively small [${\cal
O}(g^2)$] and have a finite limit as $\eta \rightarrow 0$. The
above improved solution is uniform for all conformal time, however
from this solution it is not clear what is the decay rate since
the unperturbed  solutions $(-\eta)^{\beta_\pm}$ are multiplied by
\emph{different} functions. In Minkowski space time the decay rate
can be extracted straightforwardly and unambiguously because it
describes in general an exponential relaxation of the amplitude
that multiplies the oscillatory phases. However, in an expanding
cosmology and in particular during a de Sitter stage, field modes
with wavelengths larger than the Hubble radius do not propagate,
they either grow or decay as a function of conformal (or comoving)
time, thus the concept of the decay rate requires further
examination.

As we explain in the following section, the decay of the amplitude
can be separated from a mass renormalization in an unambiguous
manner.

\subsubsection{Decay rate and mass renormalization:}

The relevant question that we must address is how to recognize a
decay of the amplitude from a mass renormalization in this
expression. We write the mass as the renormalized mass plus mass
renormalization counterterms in the usual form
\be M^2= M^2_R+g^2
\; \delta M^2_1 +\mathcal{O}(g^4) \; . \label{massren}
\ee
Such renormalization results in a renormalization of $\nu_{\chi}$
and of the exponents $\beta_\pm$, namely
\bea \nu_{\chi} & = &
\nu_{\chi,R} - \frac{g^2  \; \delta M^2_1}{2 \; \nu_{\chi,R} \;
H^2}~~;~~ \nu_{\chi_R} \equiv  \sqrt{\frac{9}{4}-
\frac{M^2_R}{H^2} }\label{nuren}\\
 \beta_{\pm} & =  & \beta_{\pm,R} \mp \frac{g^2  \;
\delta M^2_1}{2 \; \nu_{\chi,R} \; H^2}~~;~~\beta_{\pm,R}=
\frac{1}{2}\pm \nu_{\chi_R}\label{betasren}  \; .
\eea
We now insert $\beta_{\pm}$ as given Eq.(\ref{betasren}) in
Eq.(\ref{solX0}) and to order $ g^2 $ we find
\bea\label{solX0R}
&&X_{\vec{0}}(\eta) = a \; (-\eta)^{\beta_+}\left(1-\frac{g^2}{2
\; (2 \; \pi \; H)^2 \; \nu_{\chi}} \left\{ \left[ \log\epsilon +
\gamma +  \psi(\beta_-) +(2\; \pi)^2 \;  \delta M^2_1 \right] \log
\frac{\eta}{\eta_0} + \mbox{non-secular}\right\} \right) +\cr \cr
&&+b \; (-\eta)^{\beta_-}\left(1+\frac{g^2}{2 \; (2 \; \pi \; H)^2
\; \nu_{\chi}} \left\{ \left[ \log\epsilon + \gamma +
\psi(\beta_+) + (2\; \pi)^2 \;  \delta M^2_1 \right] \log
\frac{\eta}{\eta_0} + \mbox{non-secular}\right\} \right) \; .
\eea
In the expression above and in what follows we have suppressed the
subscript R to avoid cluttering of notation, but it should be
understood that all quantities are renormalized.

We see that mass renormalization cannot cancel both secular terms,
and that the mass renormalization correction for the growing and
decaying solutions have \emph{opposite sign} . Therefore, choosing
the mass renormalization counterterm $\delta M^2_1$ to be given by
\be \label{renofmass} \delta M^2_1 \equiv - \frac1{(2\; \pi)^2}
\left\{ \log\epsilon + \gamma + \frac12 \left[ \psi(\beta_+) +
\psi(\beta_-)\right] \right\} \ee \noindent cancels the
logarithmic short distance divergence and leaves a finite
contribution that multiplies \emph{both} solutions equally.

With this choice of mass counterterm, the solution of the equation
of motion to first order becomes
\be
X_{\vec{0}}(\eta) = a \; (-\eta)^{\beta_+}\left[1 + \Gamma_1
\; \log \frac{\eta}{\eta_0} + \mbox{non-secular}\right] + b \;
(-\eta)^{\beta_-}\left[1+ \Gamma_1 \; \log \frac{\eta}{\eta_0} +
\mbox{non-secular}\right] \; .
\ee
\noindent where
\be \Gamma_1 = \frac{g^2 \; \tan\pi\nu_{\chi}}{16 \; \pi \;
\nu_{\chi}
 \; H^2}\; .
\label{gamma}
\ee
and we used the relation
$$
\psi\left(\frac12 + \nu\right) - \psi\left(\frac12 - \nu\right)
= \pi \; \tan[\pi \; \nu]
$$
We can now apply the DRG which exponentiates the secular terms and
gives as the DRG-improved solution after mass renormalization \be
X_{\vec{0}}(\eta)=
\left[\frac{\eta}{\tilde{\eta}_0}\right]^{\Gamma_1}\Bigg\{a(\tilde{\eta}_0)
\; (-\eta)^{\beta_+} [1+\mathcal{O}(g^2)]+b(\tilde{\eta}_0) \;
(-\eta)^{\beta_-}[1+ \mathcal{O}(g^2)]\Bigg\}\;.\label{resumX0ren}
\ee We now clearly see that a decay rate $\Gamma_1$ can be
unambiguously identified from the contribution that multiplies
\emph{both} solutions, whereas the mass renormalization enters
with different signs for each solution.

Since the amplitudes $a(\tilde{\eta}_0),b(\tilde{\eta}_0)$ obey
the dynamical renormalization group equations
(\ref{DRGA},\ref{DRGB}), the DRG improved solution
(\ref{resumX0ren}) is \emph{independent} of the scale
$\tilde{\eta}_0$, a change in this scale is compensated by a
change in the amplitudes.

The decay rate in de Sitter space-time can be read off from the
above expression since by setting for simplicity
$\tilde{\eta}_0\equiv \eta_0$ \be
\left[\frac{\eta}{\eta_0}\right]^{\Gamma_1} = e^{-\Gamma_1 \; H \;
t} \equiv e^{-\Gamma_{dS} \; t}. \label{GammaDS} \ee and $
\Gamma_{dS} = H \; \Gamma_1 $.

Eq.(\ref{gamma}) can be analytically continued for $ M > \frac32 \; H $ as
\be\label{Mgra}
\Gamma_1 =   \frac{g^2 \; \tanh\left[\pi\sqrt{\frac{M^2}{H^2} -
\frac{9}{4}}\right]}{16 \; \pi \;  H^2 \; \sqrt{\frac{M^2}{H^2} -
\frac{9}{4}}}\; .
\ee
This result agrees with that obtained by a different method in
ref.\cite{prem}.

\subsubsection{Minkowski space-time limit.}

 Minkowski space time is recovered in the limit $H\rightarrow 0$.
In such limit we find from
  eqs. (\ref{betapm}) and (\ref{Mgra})
 \bea
&&\nu_{\chi}\buildrel{H \rightarrow 0} \over= i \; \frac{M}{H} \cr \cr
 &&\Gamma_1 \buildrel{H \rightarrow 0} \over=\frac{g^2}{16\pi M H}\,\tanh\left[
\frac{\pi\,M}{H}\right] \buildrel{H \rightarrow 0} \over=
\frac{g^2}{16\pi M H}\nonumber \\
&&\beta_{\pm} \buildrel{H \rightarrow 0} \over= \pm i \;
\frac{M}{H}  \; . \eea Therefore, we find in this limit \be
\left[\frac{\eta}{\eta_0}\right]^{\Gamma_1} = e^{-\frac{g^2}{16\pi
M }t} = e^{-\Gamma_M t}\; , \label{limitzeromom}   \ee \noindent
which displays the exponential decay of the amplitude with the
correct decay rate $\Gamma_{M}= g^2/(16 \pi M)$ for
long-wavelength excitations with mass $M$ in Minkowski space-time.
Since $-\eta = e^{-Ht}/H$  up to an overall normalization of the
field the dynamical renormalization group improved solution  in
this limit is given by \be X_0(t) = e^{-\frac{g^2}{16\pi M
}t}\left[A \; e^{i M t}+ B \; e^{-iMt}\right] \ee \noindent which
is the correct solution for a zero momentum excitation in
Minkowski space-time.

Thus, clearly the dynamical renormalization group provides a
consistent resummation program to study relaxation in a
cosmological setting.

This comparison highlights that the decay rate in de Sitter space
time is related to that of Minkowski space time as \be
\Gamma_{dS}\equiv \Gamma_1 \;  H = \Gamma_{M} \;
\frac{M}{\nu_{\chi} \; H} \;  \tan[\pi \nu_{\chi}] \; .
\label{ratecompa}
\ee
For $M \ll H$ (for the inflaton this corresponds to the slow-roll
limit) we find that the decay rate during inflation is
\emph{enhanced} as compared to that in Minkowski space-time \be
\frac{\Gamma_{dS}}{\Gamma_{M}} \buildrel{M \ll H} \over= \frac{2
\; H}{\pi \;  M} \gg 1 \; . \label{enha}
\ee
The decay rate $\Gamma_{dS}=H \Gamma_1$ with $\Gamma_1$ given by
the result (\ref{Mgra}) has a noteworthy interpretation in terms
of the Hawking temperature associated with the horizon in de
Sitter space-time $T_H = \frac{H}{2\pi}$\cite{birr} which can be
seen as follows.

The solution of the free equation of motion for the zero mode 
given by Eq.(\ref{X0}) in comoving time  is given by 
\be 
X_{\vec{0}}(t)= A \; e^{i\omega_+ t} + B \; e^{i\omega_-t}~~;~~ 
\omega_{\pm} =\frac{3 \; i  \; H}{2}\pm \sqrt{M^2-\frac{9}{4} \; H^2}\;. 
\ee
The homogeneous modes are propagating for $M>3H/2$. 
As compared to the case in Minkowski
space-time we can identify $\omega^{\pm}$ as the complex poles
that determine the free field evolution. In terms of these complex
poles it is straightforward to see that for a propagating mode
($M>3H/2$)  $\Gamma_{dS}$ can be written as follows. 
\be
\Gamma_{dS} = \frac{g^2}{16\pi} \;
\frac{1+2N(\omega_+)}{\mathrm{Re}(\omega_+)}\;,\label{temp} \ee
\noindent with $N(\omega)$ the Bose-Einstein distribution function
at the Hawking temperature, namely \be N(\omega)=
\frac{1}{e^{\frac{\omega}{T_H}}-1} \; . 
\ee The expression
(\ref{temp}) is similar to the expression for the decay rate in
Minkowski space time at finite temperature and to lowest order in
the coupling in terms of the pole frequency for the free
field\cite{bdv96}.

Thus at least in the case in which there is propagation, namely
$M>3H/2$,  the decay rate can be identified as being that of
Minkowski space time at the Hawking temperature.  This noteworthy
property of the decay rate for superhorizon modes has also been
discussed in ref.\cite{prem}.

\subsection{ Modes that cross the horizon during inflation
$ \vec{k} \neq 0~~,~~ \eta \rightarrow 0^-$ }\label{gener}

For arbitrary $k$ our integral equation (\ref{eqnofmot}) takes the
form \be \label{kgen}
X''_{\vk}(\eta)+\left[k^2-\frac{1}{\eta^2}\Big(\nu^2_{\chi}-
\frac{1}{4} \Big) \right] X_{\vk}(\eta)- \left(\frac{g}{2 \; \pi
\; H}\right)^2 \; \frac1{\eta} \; \int_{\eta_0}^{\eta}
\frac{d\eta'}{\eta'} \; \frac{(\eta-\eta') \; \cos k(\eta-\eta')}{
(\eta-\eta')^2 + ( \epsilon \; \eta')^2 } \; X_{\vk}(\eta')=0 \ee
where we used eqs.(\ref{kernelconf} )-(\ref{PP}).

To first order in $ g^2 $ the solution $ X_{1,\vk}(\eta) $ given
by Eq.(\ref{forsol}) becomes \be\label{x1kg}
 X_{1,\vk}(\eta) = \frac{1}{(2\pi H)^2} \int_{\eta_0}^{\eta}
\frac{d\eta'}{\eta'} \;  \mathcal{G}(k;\eta,\eta')
\int_{\eta_0}^{\eta'}\frac{d\eta''}{\eta''} \;
\frac{(\eta'-\eta'') \; \cos k(\eta'-\eta'')}{ (\eta'-\eta'')^2 +
( \epsilon \; \eta'')^2 } \; X_{0,\vk}(\eta'') \;, \ee where
$\mathcal{G}(k;\eta,\eta')$ is given by Eq.(\ref{GF}) and for
simplicity we consider the solution
$$
 X_{0,\vk}(\eta) = A_{\vk}\; g_{\nu_{\chi}}(k;\eta) \; .
$$
with $g_{\nu_{\chi}}(k;\eta)$ the mode function with Bunch-Davies
initial condition.  Eq.(\ref{x1kg}) can be written in the
following form \be \label{x1u}
 X_{1,\vk}(\eta) = A_{\vk} \int_{\eta_0}^{\eta} \frac{d\eta'}{\eta'} \;
 g_{\nu_{\chi}}(k;\eta') \; J(\eta,\eta')  \; ,
\ee where \be \label{jota} J(\eta,\eta') =
\frac{\sqrt{\eta}}{8\pi^2 H^2}
 \; \int_{\eta'}^{\eta} \frac{d\eta''}{\sqrt\eta''}
\frac{(\eta'-\eta'') \; \cos k(\eta'-\eta'')}{ (\eta'-\eta'')^2 +
( \epsilon \; \eta')^2 } \; \mbox{Im}\left[
H^{(1)}_{\nu_{\chi}}(k\eta) \; H^{(2)}_{\nu_{\chi}}(k\eta'')
\right]\;. \ee

Our goal is to evaluate $  X_{1,\vk}(\eta) $ for $ \eta \to 0^- $.
In order to achieve this we need $ g_{\nu_{\chi}}(k;\eta) $ and
the integrand in Eq.(\ref{jota}) for small arguments:
\bea\label{gHH} && g_{\nu}(k;\eta) \buildrel{\eta\rightarrow 0^-
}\over= \frac{\sqrt{\pi \; \eta}}{2} \; i^{-\nu-\frac{1}{2}}
\left\{ \frac{i \; \Gamma(\nu)}{\pi} \; \left( \frac{2}{k \; \eta}
\right)^{\nu} \left[1 + {\cal O}\left(k^2 \; \eta^2 \right)
\right]
+ \frac{1 - i \; \cot\pi\nu }{ \Gamma(\nu+1)} \left( \frac{k \;
\eta}{2}\right)^{\nu}\left[1 + {\cal O}\left(k^2 \; \eta^2 \right)
\right]\right\} \; , \cr \cr &&\mbox{Im}\left[
H^{(1)}_{\nu_{\chi}}(k\eta) \; H^{(2)}_{\nu_{\chi}}(k\eta')\right]
\buildrel{\eta,\eta'\rightarrow 0^- }\over=
 \frac{1}{\pi \; \nu} \left[ \left( \frac{\eta}{\eta'} \right)^{\nu} -
\left( \frac{\eta'}{\eta} \right)^{\nu}\right] \left[1 + {\cal
O}\left(\eta^2 , {\eta'}^2 \right) \right]\; . \eea Inserting
Eq.(\ref{gHH}) in Eq.(\ref{x1u}) and (\ref{jota}) yields, \be
\label{x1m}
 X_{1,\vk}(\eta)\buildrel{\eta \rightarrow 0^-} \over= A_{\vk}\;
 \frac{2^{\nu-2} \; \Gamma(\nu)}{\sqrt{\pi} \; \nu
\;  i^{\nu+\frac{3}{2}} \; k^{\nu}} \; \eta^{\frac12-\nu} \;
S\left( \frac{\eta_0}{\eta}\right) \; , \ee where \be
S\left(\frac{\eta_0}{\eta}\right) \buildrel{\eta \rightarrow 0^-}
\over= - \frac12 \int_{1}^{\frac{\eta_0}{\eta}} \frac{dy}{y}
\left(y^{-2\,\nu} - 1 \right) \int_{1}^{\frac{\eta_0}{y \, \eta}}
\frac{1}{(1+i\epsilon)t - 1} \; \frac{dt}{t^{\nu+\frac12}} +
(\epsilon \to - \epsilon) \; .
\ee
Carrying out the integrations leads to the following result
\be
\eta^{\frac12-\nu} \; S\left(\frac{\eta_0}{\eta}\right)
\buildrel{\eta \rightarrow 0^-} \over=\eta^{\frac12-\nu}\left\{
\left[ \psi\left(\nu + \frac12 \right) + \gamma + \ln\epsilon
\right] \log\frac{\eta}{\eta_0} + {\cal
O}\left(\eta_0\right)\right\} \; .
\ee
Notice that this $ \eta
\rightarrow 0^- $ behavior turns out to be $k$-independent. This
is due to the fact that the term $k^2$ in Eq.(\ref{kgen}) becomes
negligible compared to the $ \frac1{\eta^2} $ term for  $ \eta
\rightarrow 0^- $ after the modes cross the horizon.

After mass renormalization according to Eq.(\ref{renofmass}) the
logarithmic short distance singularity $\ln\epsilon$ is cancelled and we 
find for the
mode functions the following result up to order $ g^2 $,
\be
 X_{0,\vk}(\eta) + g^2 \;  X_{1,\vk}(\eta)
\buildrel{\eta \rightarrow 0^-}\over= A_{\vk}\;
 \frac{2^{\nu-1} \; \Gamma(\nu)}{\sqrt{\pi}
\;  i^{\nu-\frac{1}{2}} \; k^{\nu}} \; \eta^{\frac12-\nu} \left[ 1
+ \frac{g^2 \; \tan\pi\nu_{\chi}}{16 \; \pi \; \nu_{\chi}
 \; H^2} \; \log\frac{\eta}{\eta_0} + {\cal O}\left(\eta_0\right)\right]
\; . \ee This result features  the secular term $ \log \eta $
which is resummed by implementing the DRG as in sec.IV-A with the
result,
\be
 X_{\vk}(\eta)\buildrel{\eta \rightarrow 0^-}\over=
\frac{2^{\nu-1} \; \Gamma(\nu)\;{\eta}^{\frac12-\nu}}{\sqrt{\pi}
\; i^{\nu-\frac{1}{2}} \; k^{\nu}} \; A_{\vk}(\tilde{\eta}_0)
\left[\frac{\eta}{\tilde{\eta}_0}\right]^{\Gamma_1}\left[1 +
 {\cal O}\left(g^2\right)\right]\;,
\ee
\noindent  where $\Gamma_1$ was defined in Eq.(\ref{gamma}). It is
clear from this result that the  behaviour for  $ \eta \rightarrow
0^- $ and  $k \neq 0$ with $|k\eta| \rightarrow 0$ is the same as
that for the case  $k=0$ [see Eq.(\ref{resumX0ren})]. This due to
the fact that the physical wavenumbers $ k \eta $ become so small
for $ \eta \rightarrow 0^- $ that they bear no relevance on the
late time dynamics. While this result could be expected on physical
grounds, it is important to see it emerge from the systematic
implementation of the DRG method.

\subsection{Wavelengths much smaller than the Hubble radius:
$|k\eta| \gg 1$:}\label{smallwave}
 Anticipating mass renormalization we write the mass in the equation of
 motion  (\ref{eqnofmot}) using Eq. (\ref{renofmass}). Furthermore for
$|k\eta| \gg 1$ corresponding to modes with
 wavelengths much smaller than the Hubble radius during inflation,
the hierarchy of equations  of motion up to $\mathcal{O}(g^2)$ is given by
\begin{eqnarray}\label{perteqnKgran}
  X''_{0,\vk}(\eta)+k^2\,X_{0,\vk}(\eta) & =  &  0  \\
   X''_{1,\vk}(\eta)+k^2\,X_{1,\vk}(\eta)  & =  &
   \mathcal{R}_1(k;\eta)  \; ,
\end{eqnarray}
   \noindent with the inhomogeneity now given by
\begin{equation}\label{inho}
  \mathcal{R}_1(k;\eta) =  -\frac{\delta M^2_1}{H^2\eta^2} \;
X_{0,\vk}(\eta)  - 2 \;
C(\eta)\int_{\eta_0}^{\eta} d\eta'~C(\eta') \;
\mathcal{K}_k(\eta,\eta') \; X_{0,\vk}(\eta') \; .
\end{equation}
where $ \mathcal{K}_k(\eta,\eta') $ is given by Eq.(\ref{kernelconf} ).

The solution of the zeroth order equation is \be X_{0,\vk}(\eta)=
A_k\, e^{-ik\eta}+B_k \, e^{ik\eta}  \; . \ee The counterterm
$\delta M^2_1$ is chosen to cancel the short distance divergence
proportional to $\ln\epsilon/\eta^2$. After straightforward but
lengthy algebra we find the following expression for the
inhomogeneity in the limit $|k\eta_0| \gg |k\eta| \gg 1$ \be
\mathcal{R}_1(k;\eta)= \frac{1}{8\pi^2 \eta^2 H^2} \Bigg\{
A_k\,e^{-ik\eta} \left[\ln \frac{\eta}{\eta_0} + i \;
\frac{\pi}{2} \right] + B_k\, e^{ik\eta} \left[\ln
\frac{\eta}{\eta_0} - i \; \frac{\pi}{2} \right] +
\cdots\Bigg\}\;, \ee \noindent where the dots stand for terms that
are subleading in the limit $ |k\eta| \gg 1$. The inhomogeneous
equation for the first order correction can now be solved in terms
of the retarded Green's function \be \mathcal{G}(\eta,\eta') =
\frac{1}{k} \; \sin[k(\eta-\eta')] \; \theta(\eta-\eta')\;.\ee

To leading order in the limit $|k\eta_0| \gg |k\eta| \gg 1$ we
find \be X_{1,\vk}(\eta) = - \frac{A_k \; e^{-ik\eta}}{32 \; \pi
\;  H \; k }\left\{ C(\eta)-C(\eta_0) - \frac{2 \; i}{\pi \;  H \;
\eta} \ln\frac{\eta}{\eta_0} +\cdots  \right\}- \frac{B_k \;
e^{ik\eta}}{32 \; \pi \;  H \;  k }\left\{ C(\eta)-C(\eta_0) +
\frac{2 \; i}{\pi \;  H \; \eta} \ln\frac{\eta}{\eta_0} +\cdots
\right\} \label{X1largek} \ee \noindent where again the dots stand
for terms that are subleading in the $|k\eta_0| \gg |k\eta| \gg 1$
limit and $C(\eta)=-1/H\eta$ is the scale factor. The term
$C(\eta)-C(\eta_0)$ in the above expression is truly a
\emph{secular} term, since it grows by a factor larger than
$e^{60}$ during inflation. The validity of the perturbative
expansion for this term is determined by the requirement that
$|k/HC(\eta)|= |k\eta| \gg1$, namely that the wavelengths are much
smaller than the Hubble radius all throughout inflation.

Thus the solution up to this order is given by
\bea
&&X_{\vk}(\eta) =   A_k\, e^{-ik\eta}\left\{ 1- \frac{g^2}{32\pi H
k}\left[ C(\eta)-C(\eta_0) \right] + i \; \frac{g^2}{16\pi^2 H^2}
\frac{\ln\frac{\eta}{\eta_0} }{k\eta}+\cdots\right\}+\nonumber \\
& & +B_k\,e^{ik\eta}\left\{ 1- \frac{g^2}{32\pi H k}\left[
C(\eta)-C(\eta_0) \right] - i \; \frac{g^2}{16\pi^2 H^2}
\frac{\ln\frac{\eta}{\eta_0}}{k\eta}+\cdots\right\} \; , \eea
\noindent where the dots stand for terms that are of higher order
in $g^2$ and subleading in the limit $|k\eta_0|\gg |k\eta| \gg 1$.
The dynamical renormalization group resummation
(\ref{RGsol}-\ref{Bamp}) leads to the following DRG improved
solution \be\label{kgransolRG} X_{\vk}(\eta)= e^{-\frac{g^2}{32\pi
H k}\left[ C(\eta)-C(\eta_0) \right]}\Bigg\{
A_k\,e^{-i[k\eta+\varphi_k(\eta)]}\left[1+\mathcal{O}(g^4)\right]+
B_k\,e^{i[k\eta+\varphi_k(\eta)]}\left[1+\mathcal{O}(g^4)\right]\Bigg\}\;,
\ee \noindent where $\varphi_k(\eta)$ is a logarithmic  phase that
is not relevant for the  decay of the amplitude, and the terms in
the brackets are truly perturbative in the long time limit for
wavelengths much smaller than the Hubble radius. In
(\ref{kgransolRG}) we have chosen the renormalization scale
$\tilde{\eta}_0$ to coincide with $\eta_0$. The DRG improved
solution (\ref{kgransolRG}) reveals the decay of the amplitude
with the scale factor. The result above has the correct limit in
Minkowski space-time as can be seen from the following argument.
In comoving time, the difference $ C(\eta)-C(\eta_0)= e^{Ht} $
therefore  in the limit $H\rightarrow 0$ \be \frac{g^2}{32\pi H
k}\left[ C(\eta)-C(\eta_0) \right] \buildrel{H \to 0} \over=
\frac{g^2}{32\pi  k} t \;,\label{Kgranlim} \ee \noindent which
gives the correct exponential relaxation of the amplitude of the
field for large momentum in Minkowski space-time as shown in the
appendix.

The results for the decay laws reproduce the decay rates in
Minkowski space time in the limit $H\rightarrow 0$ [see Eqs.
(\ref{limitzeromom}) and (\ref{Kgranlim})] thus confirming the
reliability of the dynamical renormalization group approach.

We can summarize the results obtained above as follows. Consider
the solution $ X_{0,\vk}(\eta) = g_{\nu}(k;\eta) $ of the unperturbed 
equation with Bunch-Davies initial conditions as given in eq.(\ref{gnu}).
The asymptotic behavior of the power spectrum (here we do not
include the $k^3$ normalization)  of the unperturbed solution for
modes deep inside the horizon $|k\eta|\gg 1$ and those well
outside the horizon $|k\eta| \rightarrow 0$ is given by
\bea |X_{0,\vk}(\eta)|^2 \buildrel{ |k\eta| \gg 1}\over   = & &
\frac{1}{2k} \nonumber \\
|X_{0,\vk}(\eta)|^2 \buildrel{ |k\eta| \rightarrow 0 }\over  = & &
\frac{2^{2\nu-2} \; \Gamma^2(\nu)\;{\eta}}{{\pi} \;
 \; (k\eta)^{2\nu}} 
\eea
Particle decay modifies the amplitude of the solution and
consequently the power spectrum, which after the DRG resummation
is given by
\bea 
|X_{\vk}(\eta)|^2 \buildrel{ |k\eta| \gg 1}\over   = & &
\frac{1}{2k}\; e^{-\frac{g^2}{16\pi
H k}\left[ C(\eta)-C(\eta_0) \right]} \label{in}\\
|X_{\vk}(\eta)|^2 \buildrel{ |k\eta| \rightarrow 0 }\over  = & &
\frac{2^{2\nu-2} \; \Gamma^2(\nu)\;{\eta}}{{\pi} \;
 \; (k\eta)^{2\nu}} \; A^2_{\vk}(\tilde{\eta_0})
\left[\frac{\eta}{\tilde{\eta_0}}\right]^{2\Gamma_1}\label{out}
\eea
\noindent where we have normalized the mode functions to
Bunch-Davies initial conditions at the beginning of inflation
$\eta=\eta_0$ in Eq.(\ref{in}). The solution for wavelengths larger
than the Hubble radius is independent of the scale
$\tilde{\eta}_0$ because the amplitude $A_{\vk}(\tilde{\eta}_0)$
obeys the DRG equation  Eq.(\ref{DRGA}). This amplitude at a given
scale $\tilde{\eta}_0$ is obtained by matching the asymptotic
forms of the DRG improved solution at a scale $\tilde{\eta}_0$.
Clearly this amplitude will depend on the decay law of modes deep
inside the horizon, which reflects a larger suppression of the
amplitude for long wavelength modes.

These results are general hence they are also valid for the decay of the 
quantum fluctuations of the inflaton field. Since the quantum
 fluctuations of the inflaton field seed scalar density
 perturbations the result obtained above leads us to
 \emph{conjecture} that the process of particle decay can lead to
 modifications of the power spectrum of superhorizon density
 perturbations which is obtained when the fluctuation freezes 
as $\eta \rightarrow  0^-$\cite{hu2}. 
The new renormalization scale $\tilde{\eta}_0$ will 
lead to violations of scale
 invariance much in the same way as in the renormalization group
 approach to deep inelastic scattering.

 Clearly in order to assess the possibility of corrections to the
 power spectrum of density perturbations from decay of quantum
 fluctuations,
 the following issues must be studied further: i) a full gauge
 invariant treatment of the self-energy corrections to the
 equations of motion for density perturbation, ii) a DRG-improved
 solution for the whole range of momenta. Such program is
 necessarily beyond the scope of this article but the results
 above are suggestive of potentially important corrections to the
 power spectrum resulting from the decay of quantum fluctuations.

\section{Conclusions and discussion}

The main goals of this article are a study of particle decay in
inflationary cosmology, and to introduce and implement a method
based on the dynamical renormalization group that allows to
systematically obtain the relaxational dynamics of quantum fields
and the decay law in particular.

One of the main points of this work is that during inflation or
more generally during a period of very rapid cosmological
expansion, the concept of a decay rate is ill suited to describe
the relaxational dynamics or particle decay. In these cases of
relevance in Early Universe cosmology, the lack of a global
time-like Killing vector prevents the interpretation of a decay
rate as an inclusive transition probability between asymptotic
in and out states per unit time and deems unreliable the Minkowski
space-time decay rate to describe particle decay.

The method that we propose to study the relaxational dynamics and
to extract the decay law relies on the full quantum equation of
motion of the expectation value of the field in an initial state
in linear response. The quantum equations of motion are non-local
as a consequence of loop corrections which determine the
self-energy. The perturbative solution of these non-local
equations of motion feature secular terms, namely terms that grow
in time and lead to a breakdown of the naive perturbative
expansion. The dynamical renormalization group\cite{DRG} provides
a systematic resummation of the perturbative expansion that leads
to an unambiguous understanding of the decay law. The dynamical
renormalization group program has been successfully implemented
and tested in a variety of situations in Minkowski space-time (see
\cite{DRG})  and this work extends it to the case of expanding
cosmology, in particular de Sitter space-time.

After introducing the method within a familiar model of
interacting fields in inflationary cosmology we studied the
relaxational dynamics of a massive field whose quanta decay into
massless conformally coupled particles via a trilinear coupling.
This model
 allows us to present the method and highlight several
important aspects in a simpler setting. We have studied the
relaxational dynamics and the decay law in the following cases: i)
$k=0$, namely superhorizon modes, ii) fixed $k \neq 0$ and $\eta
\rightarrow 0^-$ ($|k\eta| \rightarrow 0$), namely modes that
cross the horizon during inflation and iii) modes deep within
the Hubble radius during inflation $|k\eta| >>1$.

Cases i) and ii) are found to be equivalent insofar as their
relaxational dynamics. The decay constant in this case has a
noteworthy interpretation in terms of the Hawking temperature of
de Sitter space time and the Minkowski space-time limit
$H\rightarrow 0$ reproduces the familiar decay rate of a massive
particle into massless ones. In the case of modes that are deep
inside the Hubble radius throughout inflation, we find that the
relaxation is exponential in the \emph{scale factor}, the
amplitude decays as [see Eq. (\ref{kgransolRG})]
$e^{-\frac{g^2}{32 \pi Hk}[C(\eta)-C(\eta_0)]}$. We have confirmed
that the limit $H\rightarrow 0$ reproduces the Minkowski
space-time result.

In all cases studied here we find that the expansion
\emph{enhances} the decay. In the case of superhorizon modes we
find that for $H >> M$ (with $M$ being the mass of the decaying
field) the rate constant in de Sitter space time is larger than
that in Minkowski space time by a factor $\sim H/M $.

Our  results summarized in Eqs. (\ref{in})-(\ref{out}) for the
decay law of modes deep within the horizon  as well as those that
are superhorizon during inflation lead us to suggest potential
observational implications. In an interacting theory the quantum
fluctuations of the inflaton field will decay as a consequence of
the interaction. These quantum fluctuations (in a suitable gauge
\cite{hu,hu2}) are the quantum seeds of metric perturbations. If
these fluctuations decay as a consequence of the coupling between
the inflaton and other fields (such a coupling is typically
assumed for a post-inflationary reheating stage) then the decay of
the amplitude both for modes deep within the horizon as well as
those that cross the horizon during inflation will result in
potential corrections to the power spectrum of density
perturbations.  In particular the amplitude of superhorizon modes
will depend on the decay law of modes inside the horizon, which
displays a larger suppression of the amplitude for modes of longer
wavelength.

Furthermore, non-linear interactions are necessarily the source of
non-gaussian correlation functions, in the case of a cubic vertex
the three point function in the Born approximation also reveals
the emergence of terms that grow in time \cite{srednicki}. This
three point correlation function is a measure of non-gaussianity,
and is clearly of interest to study further if and how the decay
of inflaton fluctuations leads to non-gaussian correlations,
perhaps by implementing the DRG as a resummation of the secular
terms.

The decay of the inflaton field into other fields as well as the
decay into \emph{its own quanta}  and the implications for the
density fluctuations will be explored in a forthcoming article
\cite{proximo}.

\begin{acknowledgments}
We thank Norma G. S\'anchez for useful and illuminating
discussions. D.B.\ thanks the US NSF for support under grant
PHY-0242134,  and the Observatoire de Paris and LERMA for
hospitality during this work.
\end{acknowledgments}

\appendix

\section{Decay rate in Minkowski space-time:}

In order to compare the results obtained for the decay in
inflationary cosmology with those more familiar in Minkowski space
time we now summarize the Minkowski case. The equations of motion
in Minkowski space time are given by\cite{DRG}
\be\label{eqnofmotMink} \ddot{X}_{\vk}(t)+\omega^2_k \;
X_{\vk}(t)+ \int_{t_0}^{t} dt'\; {K}_k(t-t') \; X_{\vk}(t')=0
~~;~~\omega^2_k=k^2+M^2\ee \noindent where the non-local kernel is
given by
\be\label{kernelMink} {K}_k(t-t') = 2 \; g^2\int
\frac{d^3q}{(2\pi)^3} \;
\frac{\sin\left[(q+|\vec{k}+\vec{q}|)(t-t')\right]}{2q|\vec{k}+\vec{q}|}
\ee This kernel can be written in terms of the spectral density
$\sigma_k(\omega)$ in the form
\be \mathcal{K}_k(t-t') =
-i\int^{+\infty}_{-\infty} d\omega \; \sigma_k(\omega) \;
e^{-i\omega(t-t')} \ee \noindent with the spectral density given
by \be \sigma_k(\omega)=  g^2 \int \frac{d^3q}{16 \; \pi^3 \;  q
\; |\vk+\vec{q}|} \left[\delta(\omega-q-|\vk+\vec{q}|)-
\delta(\omega+q+|\vk+\vec{q}|) \right]
\ee
The delta functions represent the kinematic cuts for
particles and antiparticles respectively. An analysis of the
self-energy of the decaying particle reveals that the spectral
function is related to the imaginary part of the retarded
self-energy as follows\cite{DRG}
\be
\sigma_k(\omega)=
\frac{1}{\pi } \; \mathrm{Im}\Sigma(k;\omega) \ee \noindent where
$\mathrm{Im}\Sigma(k;\omega)$ is the imaginary part of the
retarded self-energy. A straightforward calculation gives the
following result \be \sigma_k(\omega) = \frac{g^2}{8\pi^2} \;
\mathrm{sign}(\omega) \; \Theta(|\omega|-k) \label{sigma} \ee The
decay rate $\Gamma_k$ is given by \be\label{decayrate} \Gamma_k =
\frac{\mathrm{Im}\Sigma(k;\omega_k)}{2\omega_k} \ee \noindent
where $\omega_k$ is the particle pole (mass shell dispersion
relation) and leads to the result \be \label{gammaMink} \Gamma_k =
\frac{g^2 }{16 \; \pi \; \omega_k} \; \Theta(\omega_k -k) \ee The
zero momentum limit gives \be \Gamma_0 = \frac{g^2 }{16\pi M}
\ee
\noindent which is the result quoted in Eq. (\ref{limitzeromom})
for the limit $H\rightarrow 0$.

The limit $M=0$ which describes the Minkowski space-time limit of
the case of wavelengths much smaller than the Hubble radius [see
section \ref{smallwave} and Eq. (\ref{perteqnKgran})] yields,
\be
\Gamma_k = \frac{g^2 }{32\pi k}
\ee
\noindent since $\Theta(0)=1/2$. This result coincides with the
$H\rightarrow 0$ limit in Eq. (\ref{Kgranlim}). A detailed study
using the dynamical renormalization group\cite{DRG} reveals that
if the particle mass shell coincides with the origin of the
multiparticle threshold there emerges a phase that depends
logarithmically on time. Such is the case in the massless limit
and the logarithmic phase is the Minkowski space-time limit of the
correction $\varphi_k(\eta)$ in the improved solution
(\ref{kgransolRG}) (for more details see ref.\cite{DRG}).

\end{document}